\documentclass{elsart}

\usepackage{graphicx}

\usepackage{amssymb}

\begin{document}

\begin{frontmatter}



\title{High-energy electron observations by PPB-BETS flight 
in Antarctica}


\author[label1]{S.Torii}, 
\author[label2]{T.Yamagami}, 
\author[label3]{T.Tamura}, 
\author[label4]{K.Yoshida}, 
\author[label5]{H.Kitamura}, 
\author[label3]{K.Anraku}, 
\author[label6]{J.Chang}, 
\author[label7]{M.Ejiri}, 
\author[label2]{I.Iijima}, 
\author[label7]{A.Kadokura}, 
\author[label1]{K.Kasahara}, 
\author[label8]{Y.Katayose}, 
\author[label9]{T.Kobayashi}, 
\author[label10]{Y.Komori}, 
\author[label2]{Y.Matsuzaka}, 
\author[label11]{K.Mizutani}, 
\author[label12]{H.Murakami}, 
\author[label2]{M.Namiki}, 
\author[label2]{J.Nishimura}, 
\author[label2]{S.Ohta}, 
\author[label2]{Y.Saito}, 
\author[label8]{M.Shibata}, 
\author[label3]{N.Tateyama}, 
\author[label7]{H.Yamagishi}, 
\author[label3]{T.Yamashita}, 
\author[label3]{T.Yuda}

\address[label1]{Research Institute for Science and Engineering, 
Waseda University, 3-4-1 Okubo, Shinjuku-ku, Tokyo 169-8555, Japan}

\address[label2]{Institute of Space and Astronautical Science, JAXA, 3-1-1 Yoshinodai, Sagamihara 229-8510, Japan}

\address[label3]{Institute of Physics, Kanagawa University, 3-27-1 Rokkakubashi, 
Kanagawa-ku, Yokohama 221-8686, Japan}

\address[label4]{Department of Electronic Information Systems, 
Shibaura Institute of Technology, 307 Fukasaku, Minuma-ku, Saitama-shi 337-8570, Japan}

\address[label5]{National Institute of Radiological Sciences, 4-9-1 Anagawa, 
Inage-ku, Chiba-shi 263-8555, Japan}

\address[label6]{Purple Mountain Observatory, Chinese Academy of Sciences, 
2 West Beijing Road, Nanjing 210008, China }

\address[label7]{National Institute of Polar Research, 1-9-10 Kaga, 
Itabashi-ku, Tokyo 173-8515, Japan}

\address[label8]{Department of Physics, Yokohama National University, 
79-1 Tokiwadai, Hodogaya-ku, Yokohama 240-8501, Japan }

\address[label9]{Department of Physics and Mathematics, Aoyama Gakuin University, 
5-10-1 Fuchinobe, Sagamihara 229-8558, Japan }

\address[label10]{Kanagawa University of Human Services, 1-10-1 Heiseicho, 
Yokosuka 238-8522, Japan}

\address[label11]{Department of Physics, Saitama University, 255 Shimo-Okubo, 
Sakura-ku, Saitama-shi 338-8570, Japan}

\address[label12]{Department of Physics, Rikkyo University, 3-34-1 Nishi-Ikebukuro, 
Toshima-ku, Tokyo 171-8501, Japan}

\begin{abstract}
We have observed cosmic-ray electrons from 10~GeV to 800~GeV 
by a long duration balloon flight using 
Polar Patrol Balloon (PPB) in Antarctica. 
The observation was carried out for 13 days at an average altitude of 35~km 
in January 2004. 
The detector is an imaging calorimeter composed of scintillating-fiber 
belts and plastic scintillators inserted between lead plates 
with 9 radiation lengths. 
The performance of the detector has been confirmed by the CERN-SPS beam test 
and also investigated by Monte-Carlo simulations. 
New telemetry system using a commercial satellite of Iridium, 
power supply by solar batteries, and automatic level control using CPU  
have successfully been developed and operated during the flight. 
From the long duration balloon observations, 
we derived the energy spectrum of cosmic-ray electrons 
in the energy range from 100~GeV to 800~GeV. 
In addition, for the first time we derived 
the electron arrival directions above 100~GeV, 
which is consistent with the isotropic distribution. 
\end{abstract}

\end{frontmatter}


\section{Introduction}
\label{sec:introduction}

High-energy cosmic-ray electrons cannot propagate far from the sources, 
because the electrons lose rapidly energy with an energy loss rate of 
the square of energy through the synchrotron and inverse Compton process. 
Kobayashi et al. (2004) \cite{kobayashi04} suggests that 
the energy spectrum of cosmic-ray electrons 
might have a structure in the energy region over several 100GeV 
due to the discrete effect of local sources. 
This means that we can identify cosmic-ray electron sources 
from the electron spectrum over several 100GeV together 
with anisotropies in arrival direction.
In addition, 
there is a possibility that 
Weakly Interacting Massive Particles (WIMPs) annihilate directly 
into electron-positron pairs and produce 
mono-energetic electrons and positrons \cite{cheng02}. 
Although the propagation through the Galaxy would broaden the line spectrum, 
the observed electron and positron spectrum could still have a
distinctive feature. 
Since there are no other known production mechanisms that would produce 
an electron and positron peak at energies of 100~GeV - 10~TeV, 
such a distinctive feature clearly indicates the existence of 
WIMP dark matter in the Galactic halo. 
Thus, the observations of high-energy electrons bring us 
unique information about sources and propagation of cosmic rays, and 
enable us to search for WIMP dark matter.

Although the cosmic-ray electrons have been observed with 
many kinds of detectors borne on balloons, 
most observations are limited below $\sim$100 GeV. 
As an exceptional instance, the observation with the emulsion chambers (ECC) 
has achieved to detect electrons beyond 1TeV 
by the large amount of exposures of $\sim14$ days in total 
for 30 years \cite{nishimura80, kobayashi02}. 
The difficulty of observation originates from that 
the electron flux itself is very small and decreases 
with energy much more rapidly than that of protons 
because of the electro-magnetic energy loss by radiation. 
The electron flux is estimated to be 
$\sim$1\% of protons around 10GeV and less than $\sim$0.2\% of
protons above 100~GeV 
from the observed energy spectra with a power-law index of 
$-3.0$ $\sim$ $-3.3$ 
for electrons \cite{torii01,kobayashi02} and 
$-2.7$ for protons \cite{haino04}. 
Therefore, for the electron observations above 100~GeV, 
we need a long duration observation by a detector 
with an excellent capability of large geometrical factor and   
powerful background rejection.

In order to meet the requirements for the electron observation, 
we have developed a highly granulated imaging calorimeter, 
the Balloon-borne Electron Telescope with Scintillating fibers (BETS), 
that preserves the superior qualities of both electronic detectors 
and emulsion chambers. 
We have successfully observed cosmic-ray 
electrons in the energy range from 10 to 100~GeV 
at Sanriku in Japan \cite{torii01}. 
The BETS was improved to observe 
atmospheric gamma rays at mountain and balloon
altitudes for calibrating the atmospheric neutrino flux calculations 
\cite{kasahara02}. 
Furthermore, 
for observing the electrons over 100GeV by a long duration balloon
experiment, 
we have achieved the development of an advanced detector, PPB-BETS, 
flown by PPB in Antarctica. 
The PPB has a capability to realize a flight for $2-4$ weeks 
in one round of the Antarctica at an altitude of $\sim$35~km. 
The PPB project is carried out by the inter-university collaboration 
with Institute of Space and Astronautical Science (ISAS), JAXA 
and National Institute of Polar Research (NIPR) 
\cite{nishimura94,kadokura02}.

In this paper, 
we describe the instrumentation of PPB-BETS, the calibration of the detector 
with accelerator beams, the observations in Antarctica, 
and the results of our experiments.

\section{Instrumentation}
\label{sec:instrument}

%
\subsection{Detector}
\begin{figure}
 \begin{center}
  \includegraphics[width=100mm]{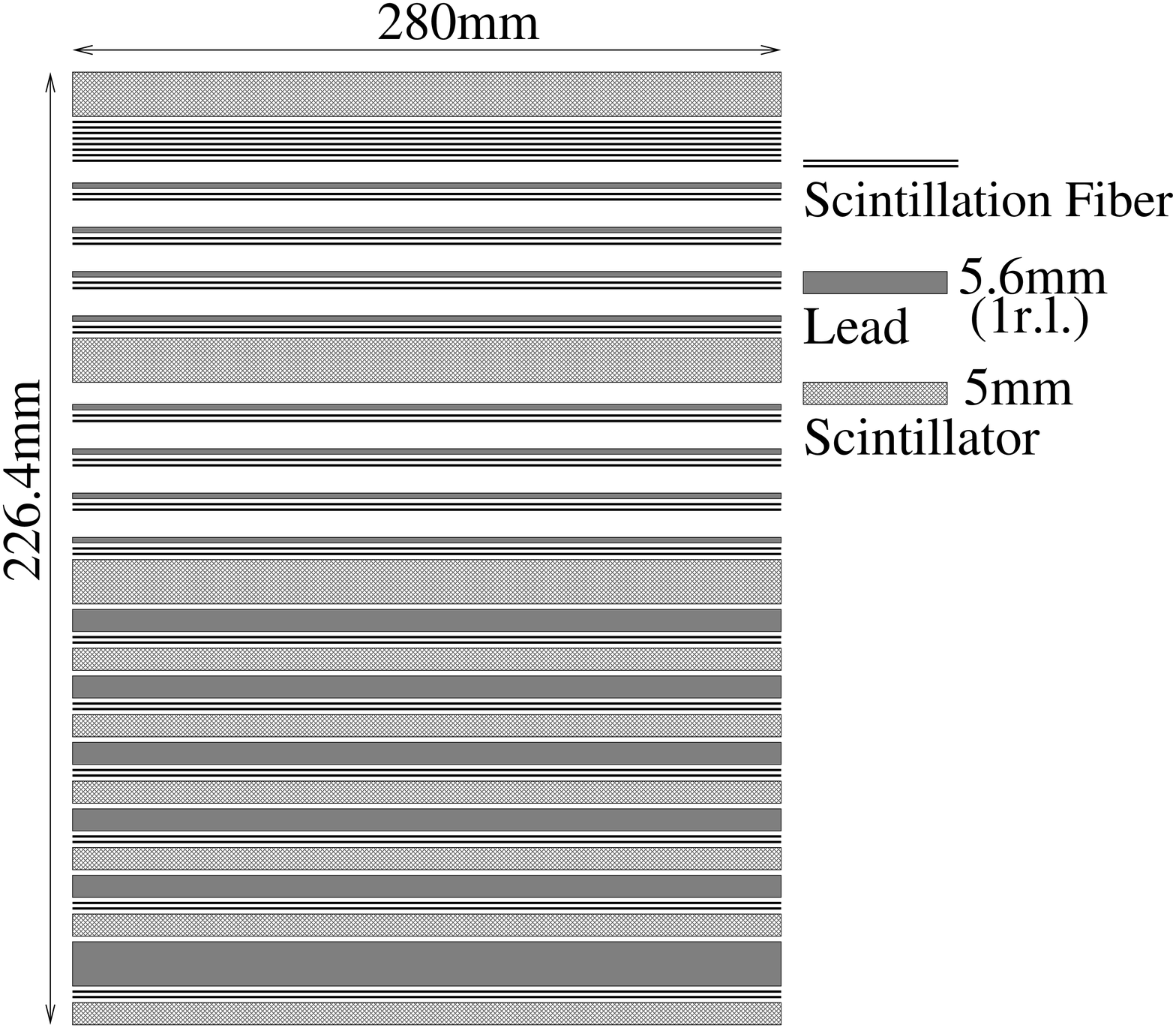}
 \end{center}
\caption{Schematic side view of the PPB-BETS detector.
See text for details.}
\label{fig:design}
\end{figure}

The PPB-BETS detector consists of 36 scintillating fiber belts, 
9 plastic scintillators, and 
14 lead plates with 9 radiation lengths (r.l.) in total. 
Each fiber belt is composed of 280 fibers 
with a 1mm square cross section for each. 
A schematic cross section of the detector is shown 
in Fig. \ref{fig:design}. 
The area of the detector is 28~cm$\times$28~cm and 
the total thickness is 23cm, 
including spacers inserted between sensitive layers. 
The scintillating fibers 
perform the function to detect the shower particles 
developing in lead plates 
and to fulfill a detailed imaging of the shower. 
Scintillating fiber belts are set in right angle 
alternately to observe the projected shower profile 
in x and y directions. 
For the read-out of scintillation light in the fibers, 
we used an image-intensified CCD camera 
in each direction of x and y. 
The plastic scintillators were adopted for the instrument trigger 
and the energy measurement. 
Thus PPB-BETS is an instrument to observe the details of 
three dimensional shower development with a timing capability.

The basic structure of PPB-BETS is similar to the BETS \cite{torii00}. 
In order to achieve the observation of 
electrons up to 1TeV by using a long duration flight, however, 
the PPB-BETS has been improved in comparison with the BETS 
at several points presented in the following. 
The thickness of lead is increased from 7 to 9 r.l.  
to observe higher energy electrons, and 
the number of plastic scintillators is increased from 3 to 9 
to detect accurately the shower development. 
The image-intensified CCD camera (I.I.-CCD) system for the read-out of
scintillating fibers has newly been developed by using a better quality 
of CCD to detect 1TeV electron showers without saturation 
of CCD signals at higher intensities of photon numbers.

The total performance of detector was approved by the beam test 
using accelerators and by the Monte-Carlo simulations. 
Monte-Carlo simulations were carried out with several codes 
such as FLUKA2002 \cite{fasso01a,fasso01b}, GEANT3 \cite{apostolakis03}, 
and EPICS \cite{kasahara03}, 
in which we could not find serious differences.

\subsection{Trigger system}

The trigger rate in actual observation is critically important 
to know the data amount transmitted by telemetry, 
since we have no guarantee of the recovery of instrument. 
Therefore, we carried out a test flight of the trigger system 
at Sanriku in advance of the flight in Antarctica. 
The trigger system is composed of 
the plastic scintillators and lead plates in same configuration 
of the PPB-BETS. 
The flight has confirmed that the trigger system 
worked well and the rate observed was consistent with 
the expectation by the simulation. 
We made also a study on the detailed performance of detector 
at CERN-SPS by using the electron beams from 10GeV to 200GeV 
and proton beams from 150 to 350GeV. 
The electron events developed purely in electro-magnetic process 
have similar and smooth shower developments. 
On the other hand, 
the proton events caused by nuclear interactions 
have the developments with much wider fluctuations 
and relatively smaller energy depositions. 
The detection efficiency of the shower trigger system was examined by
adjusting the discrimination levels at each scintillator.

\subsection{Imaging system}

For the image-intensified CCD camera, 
it is required to observe shower images 
from a single track to a shower maximum without saturation. 
The CCD (THX 7887A) has $1024{\times}1024$ pixels, and 
the signal-to-noise ratio is 70dB that means 6.3 times larger dynamic
range of CCD than the BETS of 54dB. 
Since the pixel number increases by 16 times comparing to that used 
in BETS, 
the dynamic range of CCD for the photon intensity in one fiber 
extends by two order of magnitude. 
Extrapolating the data of the CERN beam test to higher energies, 
we confirmed that the I.I.-CCD system has an enough dynamic range 
for the observation of one single track to 1TeV electrons 
with linearity. 
The one CCD image is read out with a speed of $1/30$sec by using 
12 bit ADC of 40MHz. 
Pile-up of events in CCD are taken place 
only by successive events within the timing of 10${\mu}$sec, 
and the effect can be negligible small.  
The total data processing on-board take $\sim1$sec for one event.

Since the instrument is not scheduled to be recovered, 
all of the observed data should be transfered by the Iridium 
satellite phone line and the down link to the Syowa station. 
To meet the telemetry rate of 2.4kbps by Iridium, 
the image data can be compressed on-board with a Run-Length method and 
a Huffman method. 
From the beam test, we obtained that 
the compressed image data size of one event 
is $\sim$18kByte for electrons at 100~GeV and 
$\sim$13kByte for protons at 100~GeV 
with the data compression rate of $\sim$40\%.

\subsection{Ballooning technology}

As a ballooning technology of the long duration flight, 
we have developed 
the telemetry system with the Iridium satellite phone system, 
power supply system by solar batteries, 
automatic level control system using CPU, and so on. 
Since the detector weight should be saved for loading the heavy ballast, 
we used an un-pressurized vessel with a light shield, 
designing the instrument to meet the vacuum and heat conditions 
during the long duration flight. 
We examined the validity of the instrument by environmental tests 
as long as more than one week.

We summarize the basic parameters of the PPB-BETS instrument 
in Table \ref{tab:parameters}.

\section{Observations}
\label{sec:observation}

\begin{figure}
 \begin{center}
  \includegraphics[width=100mm]{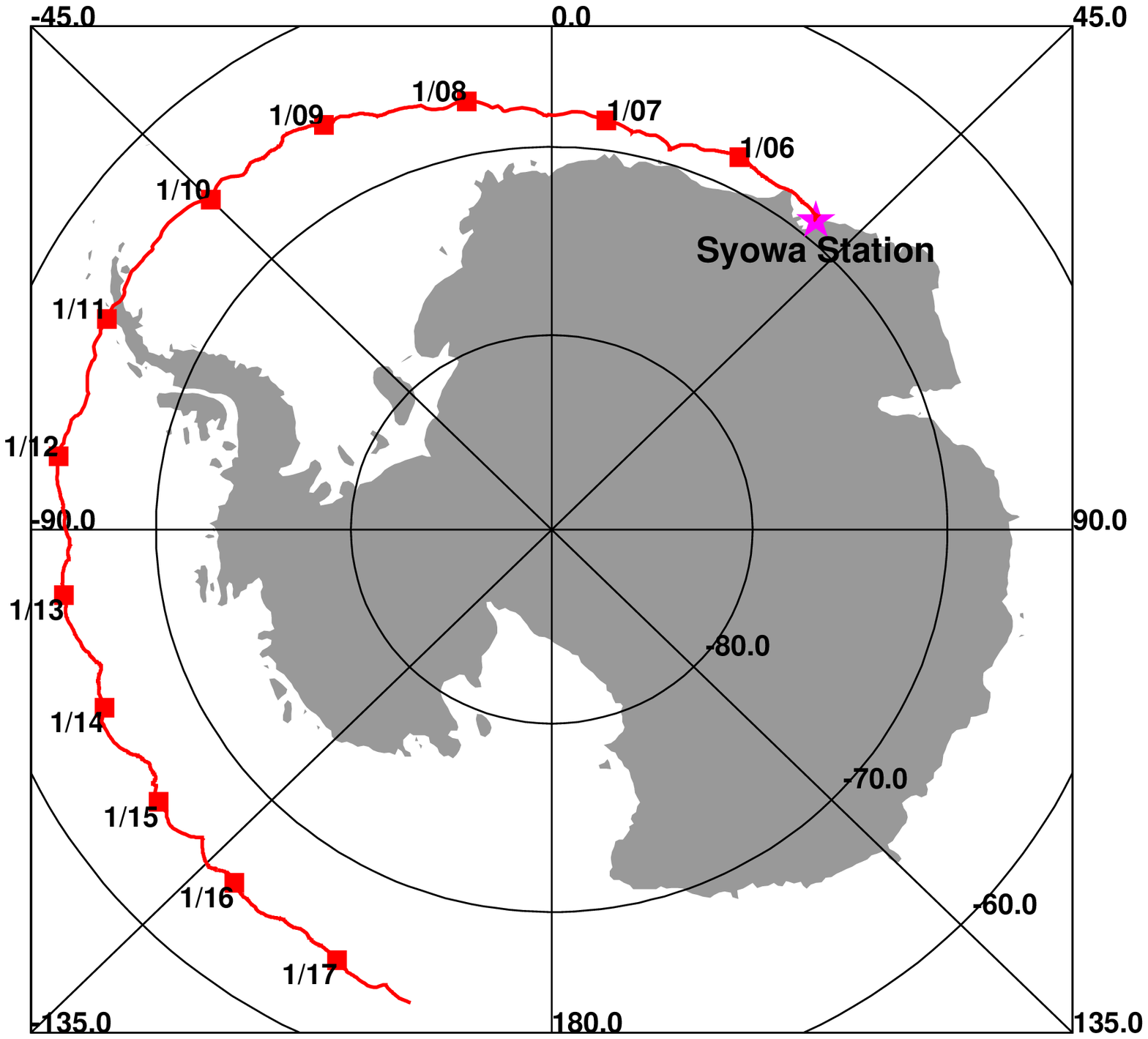}
 \end{center}
\caption{
The trajectory of the PPB-BETS in Antarctica. 
}
\label{fig:trajectory}
\end{figure}

\begin{figure}
 \begin{center}
  \includegraphics[width=100mm]{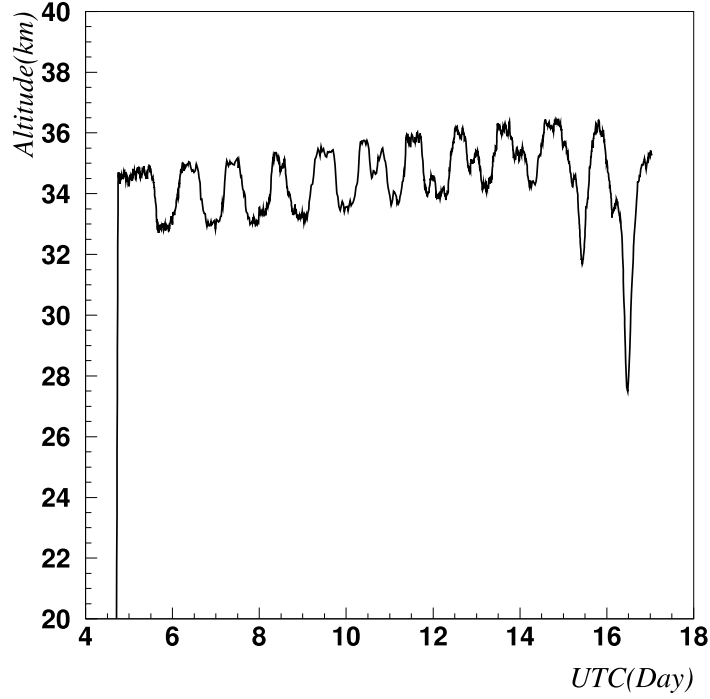}
 \end{center}
\caption{
The altitude profile of the PPB-BETS. 
}
\label{fig:altitude}
\end{figure}

The balloon was launched at the Syowa station 
(69$^{\circ}$00' latitude south, 35$^{\circ}$35' longitude east) 
in Antarctica at 15:57 on January 4, 2004 (UTC). 
The level flight was started at 18:00 on January 4, and 
continued till 1:46 on January 17, 2004 (UTC) 
at altitudes of 33-37 km (35km on average). 
The total duration of the exposure time is $2.96{\times}10^2$~hr. 
The trajectory of the PPB-BETS is presented in Fig. \ref{fig:trajectory}. 
The balloon went round the Antarctica at ${\sim}65^{\circ}$ latitude 
south from east to west with a speed of $\sim30-35$~km~h$^{-1}$. 
Power consumption of 70W in the instrument 
was normally supplied by the solar batteries. 
Automatic level control system successfully operated 
as presented in Fig. \ref{fig:altitude}.

The event trigger was executed by two modes,  
the high-energy (HE) mode and low-energy (LE) mode. 
The LE mode corresponds to electron observation over 10GeV, 
and was assigned for the observation 
during 10 hours just after launching.  
The data acquired by the LE mode were directly transfered to 
the Syowa station with a telemetry of 64kbps. 
The HE mode, 
which corresponds to the electron observation over 100GeV, 
was used through all the flight. 
The acquired data were further selected by the software trigger 
(2nd trigger) on-board, and 
transmitted to our operation room at NIPR in Japan, 
via a receiving station in US, 
with an Iridium satellite phone line at a rate of 2.4kbps. 
The commands from the operation room 
to the PPB-BETS were also sent by the Iridium line. 
The monitoring data were delivered to each institute via Internet. 
The data transfer rates were 
$\sim1{\times}10^4$ events per hour (3Hz) for the LE mode and 
$\sim70$ events per hour (0.02Hz) for the HE mode, respectively. 
The amount of data used for the current analysis 
is $\sim1/4$ of the acquired data on-board 
because of the problem of data transfer system and CCD noise.

\section{Data analysis}
\label{sec:analysis}

%
\subsection{Data sets for analysis}

\begin{figure}
 \begin{center}
  \includegraphics[width=100mm]{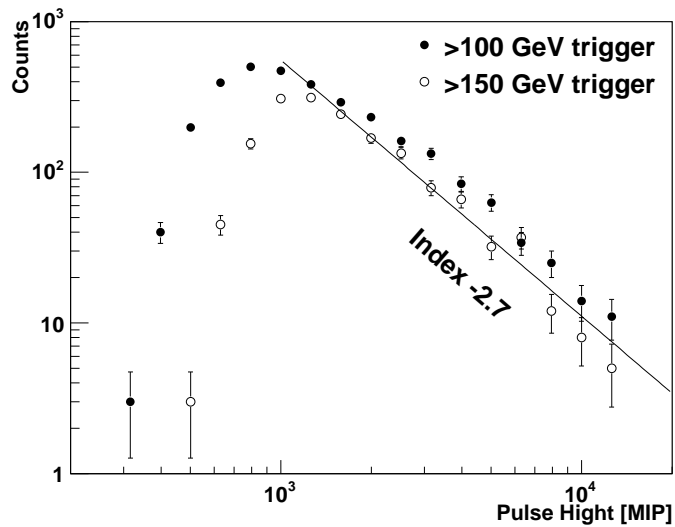}
 \end{center}
\caption{
Pulse height distributions of the plastic scintillator at 
the depth of 7 r.l. 
for the 100~GeV threshold (solid circles) and 
150~GeV threshold (open circles). 
The solid line shows a power-law spectrum with an index of $-2.7$. }
\label{fig:ph_dist}
\end{figure}

For the event triggering the HE mode, 
we adopted two main sets of discrimination levels 
of pulse heights in all these plastic scintillators. 
One set corresponds to the threshold energy of 100~GeV and 
the other to 150~GeV. 
The number of acquired events is $3.1{\times}10^3$  for the 100~GeV 
threshold and $1.6{\times}10^3$ for the 150~GeV threshold. 
There are also $1.0{\times}10^3$ events for the other thresholds 
on the HE mode. 
In this analysis, 
we used the two data sets of the 100~GeV and 150~GeV threshold 
without use of the other data including the LE mode. 
Figure~\ref{fig:ph_dist} shows the pulse height (converted to the number of 
Minimum Ionizing Particle, MIP) distribution of the plastic scintillator 
at the depth of 7 r.l. for the two thresholds. 
As shown in Fig.~\ref{fig:ph_dist}, 
the observable minimum energy is shifted properly depending on each threshold, 
and the trigger system worked normally. 
The spectra in high-energy region are compared to a power-law spectrum 
with an index of $-2.7$. 
They are consistent with the spectral shape of the cosmic-ray protons, 
since the events triggered on-board are still dominated by the protons 
that are much more abundant than the electrons.

\subsection{Energy measurement}
\begin{figure}
 \begin{center}
  \includegraphics[width=100mm]{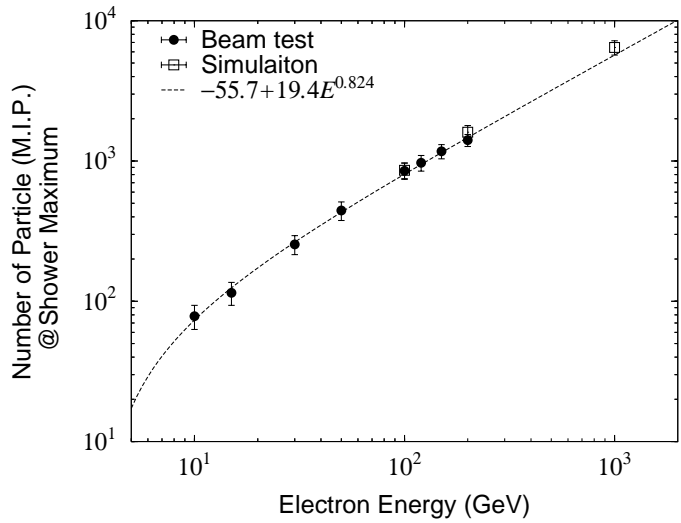}
 \end{center}
\caption{Relation of the electron energy and the number of shower particles 
at the shower maximum by the CERN-SPS beam tests with simulations. 
The incident direction is perpendicular to the detector surface. }
\label{fig:ene2mip}
\end{figure}

\begin{figure}
 \begin{center}
  \includegraphics[width=100mm]{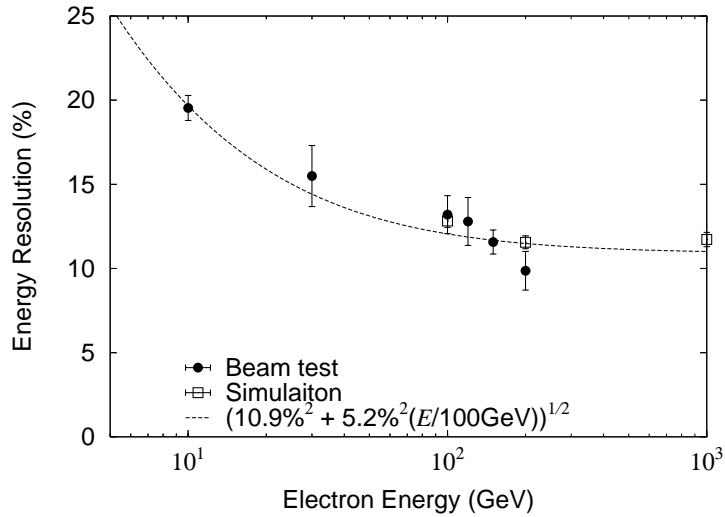}
 \end{center}
\caption{Energy dependence of the energy resolution 
for the CERN-SPS electron beams with simulations. 
The incident direction is perpendicular to the detector surface. }
\label{fig:eneres}
\end{figure}

The plastic scintillators are used for the energy measurements 
of electrons. 
Since the number of shower particles at the shower maximum is 
nearly proportional to the energy of incident electrons,  
we determined the electron energies 
by using the number of particles at the shower maximum 
in the transition curve with plastic scintillators. 
The transition curves of electro-magnetic shower 
are well represented by the following formula: 
\begin{equation}
S = N (\frac{bt}{{\cos}{\theta}})^{a-1} {\exp}(-\frac{bt}{{\cos}{\theta}}), 
\label{eq:trcv}
\end{equation}
where $S$ is the pulse heights (converted to MIP) of plastic scintillators 
at the depth of $t$~r.l., 
$N$ is the normalization factor, 
$\theta$ is the zenith angle of the shower axis, and 
$a$ and $b$ are adjustable coefficients. 
In order to derive the number of particles at the shower maximum, 
we fitted this formula to the pulse heights of the six scintillators 
at depths of 3~r.l. to 9~r.l.
As shown in Fig.~\ref{fig:ene2mip}, 
the relation of the number of particles at the shower maximum and 
the energies of electron beams at CERN-SPS is almost linear 
in the energy range of $10-200$~GeV. 
The energy resolution is 12~\% at 100~GeV as shown in Fig.~\ref{fig:eneres}, 
and consistent with the simulations.

\subsection{Angular measurement}
It is important to determine accurately a shower axis, 
since it is crucial to determine the electron energy and 
the separation of electrons from background protons and gamma rays 
as mentioned later.

\begin{figure}
 \begin{center}
  \includegraphics[width=75mm]{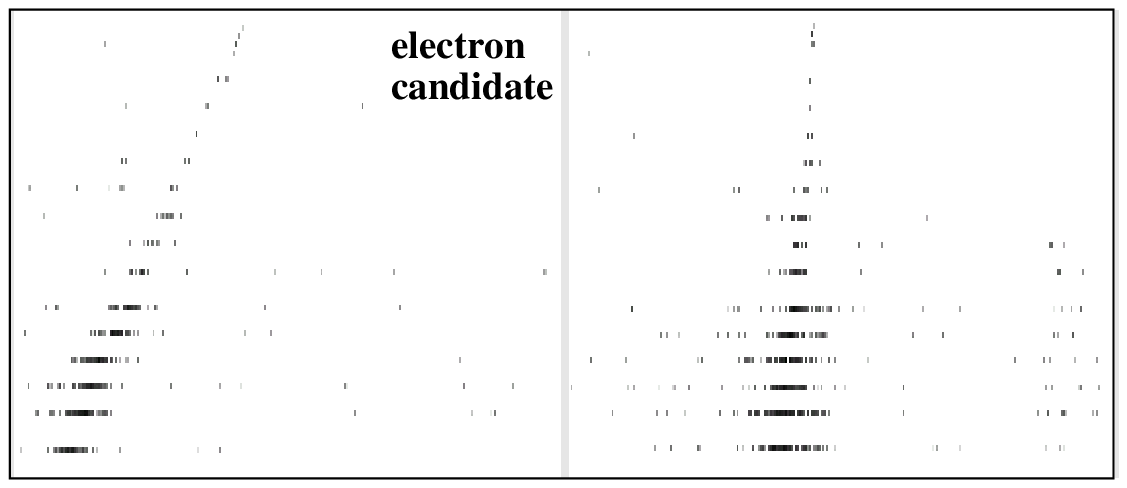}

  \includegraphics[width=75mm]{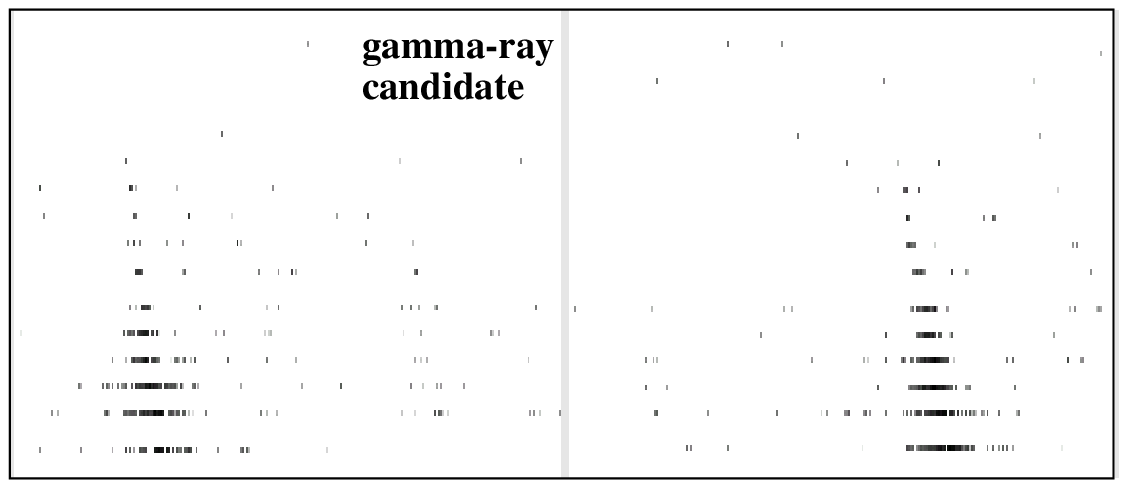}

  \includegraphics[width=75mm]{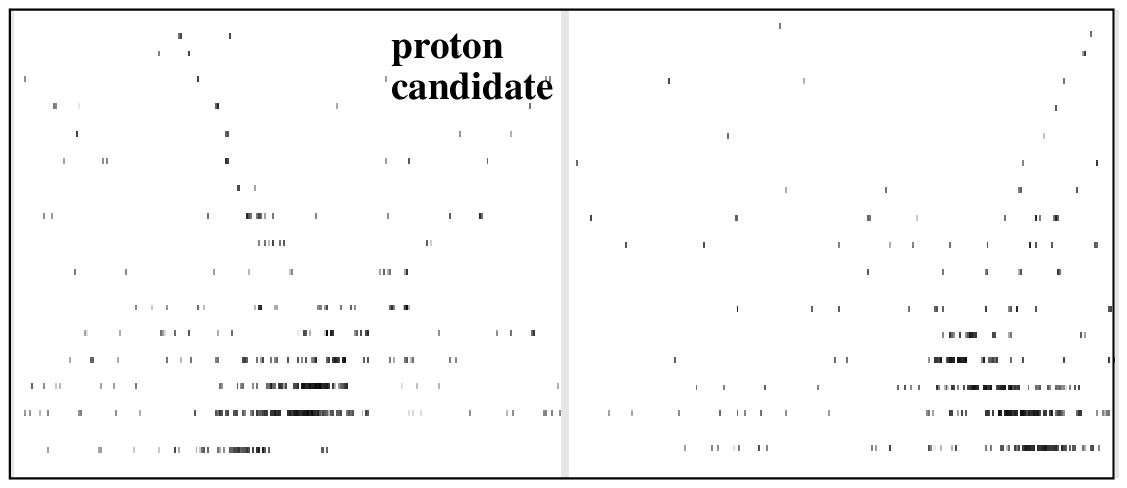}
 \end{center}
\caption{
Examples of the reconstructed images in detector space. 
Typical candidates of an electron, gamma ray, and proton 
are shown in x and y directions. 
}
\label{fig:candidates}
\end{figure}

For the acquired events in the flight, 
we reconstructed the raw CCD images of showers to 
the fiber positions in detector space by using the positions of each fiber 
on the CCD image. 
The positions of each fiber were obtained in advance of the flight 
by observing cosmic-ray muon tracks at ground level. 
The relative displacement of the fiber positions during the flight 
was calibrated by using the LED-illuminated fibers. 
Examples of the reconstructed shower image observed in the flight 
are presented in Fig. \ref{fig:candidates}.

\begin{figure}
 \begin{center}
  \includegraphics[width=100mm]{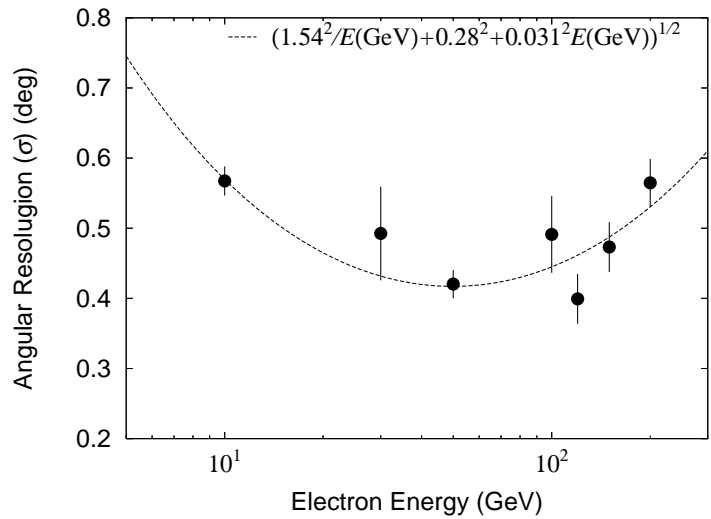}
 \end{center}
\caption{Energy dependence of the angular resolution 
for the perpendicular electron beams at CERN-SPS.}
\label{fig:beam_ang_res}
\end{figure}

From the reconstructed shower image, 
we determine the shower axis in the following. 
At first, 
below 3~r.l. in the detector, 
we derive centric positions of shower lateral spreads 
in each fiber layer. 
By fitting these centric positions with a linear least-squares method 
we determine a tentative shower axis, 
and trace back to the next upper layer. 
Adding a new centric position in the upper layer 
by using only fiber signals near the tentative shower axis ($\pm5$ fibers), 
we fit the shower axis once again. 
This process is repeated up to the top layer of 0~r.l., 
and the shower axis is determined. 
Figure \ref{fig:beam_ang_res} shows the angular resolution by the beam
test,  which ranges from 0.40 to 0.57 degree 
with electron energies of 10 to 200~GeV. 
Although they are considerably better than 
those of the BETS of $\sim1^{\circ}$, 
they are worse than the simulated angular resolution of 
$\sim0.15^{\circ}$ above 100~GeV. 
It is caused by errors of the fiber positions 
in detector space.

\subsection{Electron selection}

The electron selection from backgrounds has been carried out by the method 
used in BETS as explained below. 
At first, proton backgrounds are reduced 
with the on-board trigger system. 
The discrimination levels are set with each scintillator. 
Second, we selected the events whose shower axes pass through the top and 
bottom of the detector within a zenith angle of 30 degrees. 
This selection reduces the proton background events triggered on-board, 
which are incident from the side of the detector. 
Third, we introduced the ratio of energy deposition 
within 5mm from the shower axis to the total ($RE$ parameter) 
as described in Torii et al. (2001) \cite{torii01}. 
A typical event of electron-induced shower presents a narrower lateral spread 
concentrated along the shower axis, as shown in Fig. \ref{fig:candidates}. 
On the contrary, 
that of proton-induced shower shows a wider lateral spread. 
We characterize this physical property by the $RE$ parameter. 
The electron candidates were selected by the $RE$ distribution 
under the condition that the $RE$ values are greater than 0.75 \cite{yoshida08a}.

As for the separation between electrons and gamma rays, 
electrons could be identified by the presence of hits 
in the scintillating fibers at 0 r.l. along the shower axis. 
Incident electrons leave signals on the fibers along the shower axis. 
On the other hand, 
incident gamma rays leave no signals except for the back-scattered particles. 
We judged electrons as the events whose distances are less than or equal to 5~mm. 
Gamma-ray events are rejected by 90~\% \cite{yoshida08a}.

As a result, 
the number of the electron candidates above 100~GeV is 84 events 
among the $4.7{\times}10^3$ analyzed events in the HE mode.

\subsection{Derivation of electron energy spectrum}
From the electron events,  
with the above selection and energy determination, 
we derived the cosmic-ray electron spectrum by using the following formula: 
\begin{equation}
J_{\rm e}(E) = 
( \frac{N_{\rm e} C_{\rm RE} C_{\rm eg}}{S{\Omega}T {\Delta}E} 
C_{\rm enh} - C_{\rm 2nd} ) C_{\rm atm}, 
\label{eq:espec}
\end{equation}
where $N_{\rm e}$ is the number of electron candidates, 
$S{\Omega}$ the effective geometrical factor, 
$T$ the observed live time, 
${\Delta}E$ the energy interval, 
$C_{\rm RE}$ the correction factor of the proton contamination 
in the $RE$-cut with energy dependence, 
$C_{\rm eg}$ the correction factor of the gamma-ray contamination 
in the gamma-ray rejection with energy dependence, 
$C_{\rm enh}$ the correction of enhancement of flux 
due to the energy resolution, 
$C_{\rm 2nd}$ the flux of secondary (atmospheric) electrons 
at the observation level, and 
$C_{\rm atm}$ the correction factor of energy loss of primary electrons 
in the overlying atmosphere.

\begin{figure}
 \begin{center}
  \includegraphics[width=100mm]{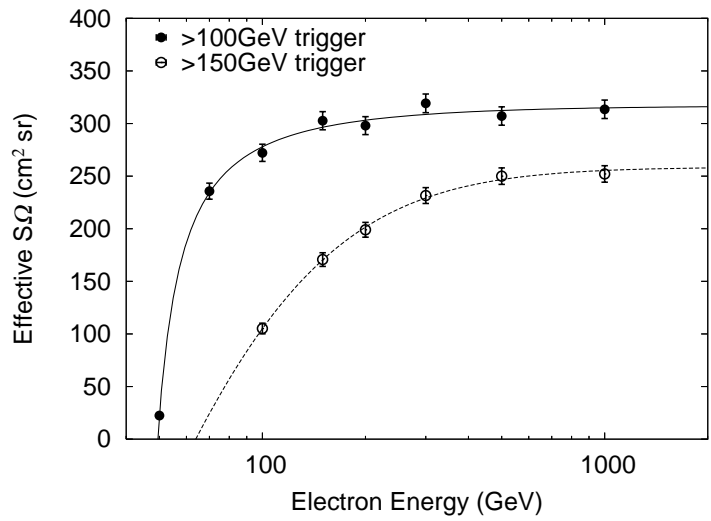}
 \end{center}
\caption{The effective geometrical factor $S{\Omega}$ with the Monte-Carlo 
simulations under the same condition of the experiment. }
\label{fig:somega}
\end{figure}

The effective geometrical factor, $S{\Omega}$, is derived 
by Monte-Carlo simulations under the same condition of the experiment. 
Figure~\ref{fig:somega} shows the effective geometrical factor with electron
energies for the 100~GeV threshold and 150~GeV threshold in the HE mode.

The uncertainty of the energy determination has the effect of enhancing the 
absolute flux of electrons, in particular, 
for a steep power-law spectrum. 
We derived the enhancement factor $C_{\rm enh}$ above 100~GeV of 0.98 
due to the energy resolution presented in Fig.~\ref{fig:eneres}.

In order to derive the primary electron spectrum, 
we subtracted the secondary electrons produced by the interactions 
of cosmic-ray nuclei. 
The energy spectrum of the estimated atmospheric electrons are 
represented by 
\begin{equation}
C_{\rm 2nd}(E) = 1.32{\times}10^{-5} (\frac{E}{\rm 100GeV})^{-2.73} 
({\rm m^{-2} s^{-1} sr^{-1} GeV^{-1}})
\end{equation}
at the altitude of 7.4~g~cm$^{-2}$ \cite{nishimura80, yoshida06}.

The atmospheric correction factor, $C_{\rm atm}$, is caused by bremsstrahlung 
energy losses of the primary electrons. 
In the case of a single power-law spectrum with an index of $-{\gamma}$, 
the energy-loss correction due to the overlying atmosphere is 
given by ${\alpha} = e^{A({\gamma}-1) t/({\gamma}-1)}$, 
where $t$ is the depth in r.l., and the notation $A$ refers 
to the formula used in electro-magnetic shower theory \cite{nishimura80}. 
The energy-loss correction, $\alpha$, 
corresponds to the correction factor of flux as follows: 
$C_{\rm atm} = {\alpha}^{\gamma-1}$. 
In the case of $\gamma=3.0$ and 7.4~g~cm$^{-2}$ depth, 
the atmospheric correction factor is 
$C_{\rm atm} = 1.37$ for ${\alpha} = 1.17$.

From the raw numbers of electron candidates, 
we evaluated the electron numbers by correcting the proton contamination 
in the $RE$ cut and the gamma-ray contamination in the gamma-ray rejection. 
Since the power-law index of the primary proton spectrum is $-2.7$, 
which is much harder than that of the electrons, 
the contamination of protons increases with energy. 
Atmospheric gamma rays are also produced by the interaction of primary protons 
with atmospheric nuclei. 
Therefore, 
the index of gamma rays is the same with that of protons. 
This brings the same energy dependence for 
the contamination of gamma rays with the protons. 
Hence, 
the correction factors are estimated to be 
$C_{\rm RE} = 1-0.325\times(E/100{\rm GeV})^{{\gamma}-2.7}$ 
for the proton contamination and 
$C_{\rm eg} = 1-0.176\times(E/100{\rm GeV})^{{\gamma}-2.7}$ 
for the gamma-ray contamination 
from the Monte-Carlo simulations.

\subsection{Check on event identification}

\begin{figure}
 \begin{center}
  \includegraphics[width=100mm]{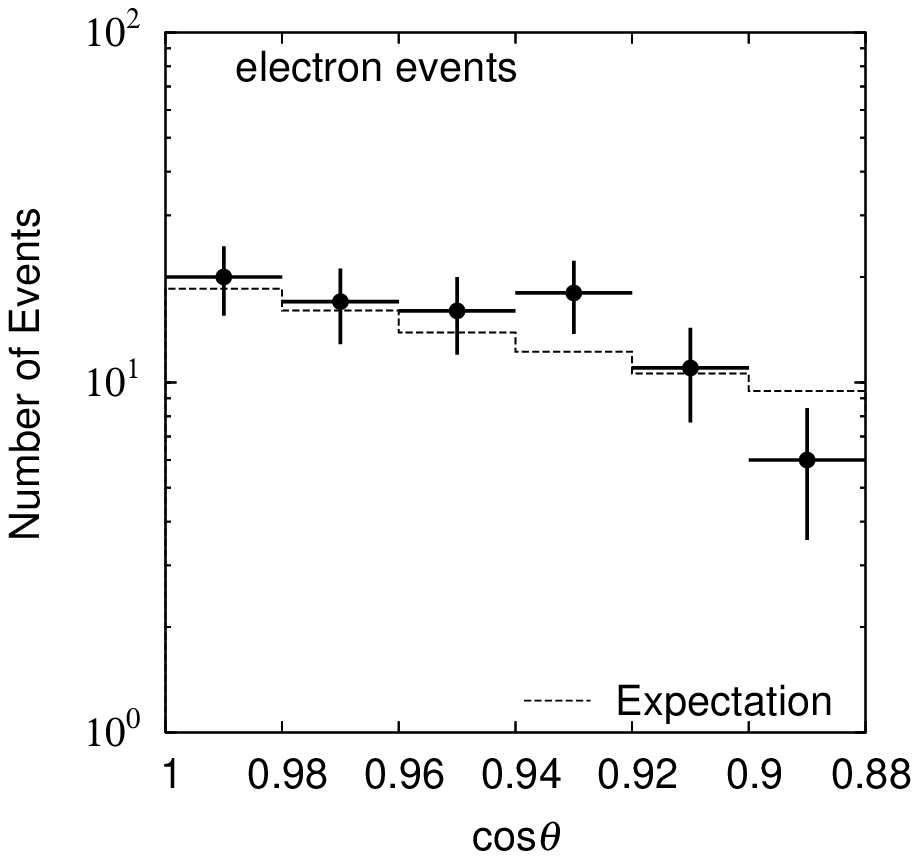}
  
  \includegraphics[width=100mm]{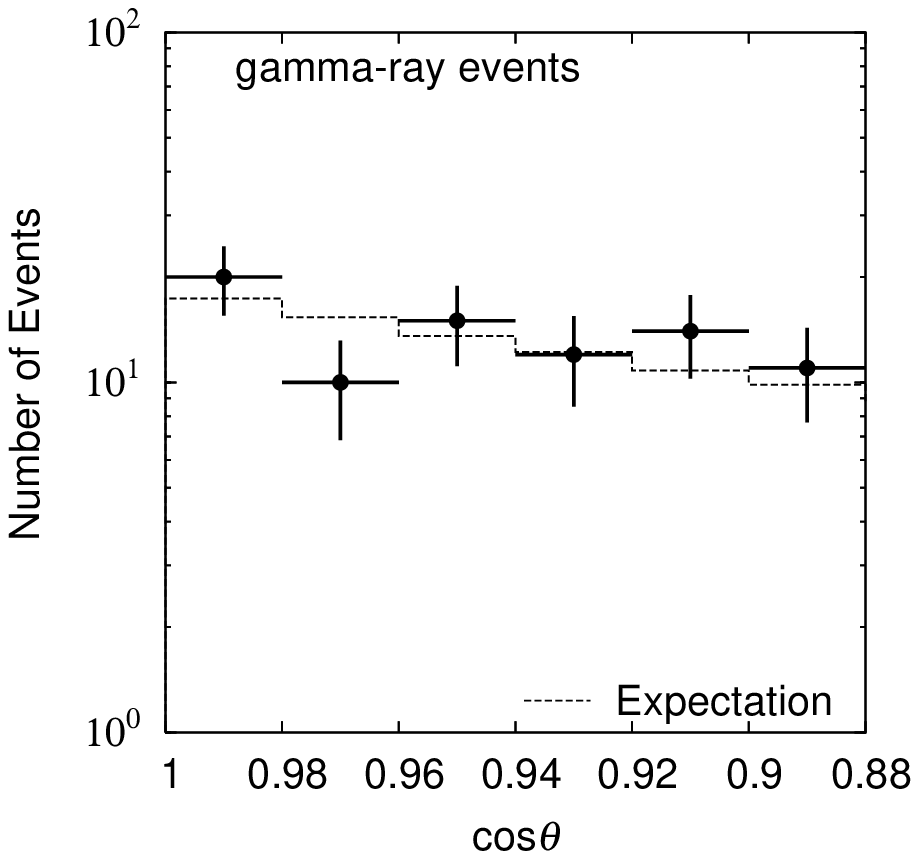}
 \end{center}
\caption{Zenith angle distributions for events classified as 
electrons and gamma rays with the expected distributions. }
\label{fig:zenith_cos}
\end{figure}

In order to check on the event identification, 
we derived the zenith angle distribution of electron and gamma-ray events.  
For cosmic-ray electrons, 
the zenith angle distribution will be isotropic, 
and thus for a plane detector the observed angular distribution 
will be proportional to ${\cos}{\theta} d({\cos}{\theta})$, 
neglecting the effect of the finite detector seize. 
The distribution for atmospheric gamma rays should be 
constant with ${\cos}{\theta}$, 
since the gamma rays are produced in the overlying residual atmosphere 
and thus have a $1/{\cos}{\theta}$ enhancement 
relative to the isotropic primary hadron flux. 
Zenith angle distributions for the electrons and gamma rays 
are presented in Fig.~\ref{fig:zenith_cos}, and 
in good agreement with the expected forms with the finite detector size.

\begin{figure}
 \begin{center}
  \includegraphics[width=100mm]{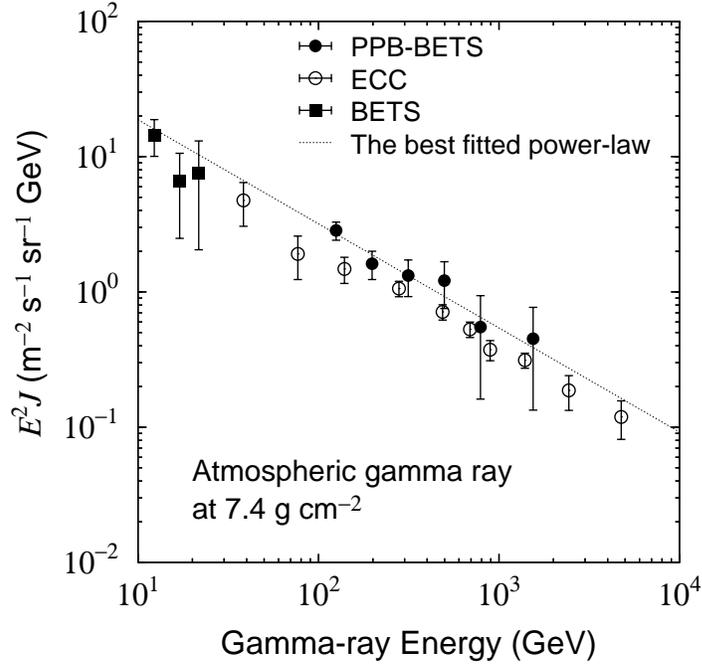}
 \end{center}
\caption{Energy spectrum of atmospheric gamma rays 
observed with PPB-BETS, 
compared to the BETS \cite{kasahara02} and ECC \cite{yoshida06}. 
The gamma-ray fluxes are normalized to 7.4 g~cm$^{-2}$ equivalent altitude.  
The dash line shows the best fit power-law function of PPB-BETS data 
with an index of $-2.77{\pm}0.21$. }
\label{fig:gspec}
\end{figure}

We also derived the vertical spectrum of atmospheric gamma rays 
at 7.4~g~cm$^{-2}$ originated from hadronic interactions, 
subtracting the bremsstrahlung gamma rays produced by primary 
cosmic-ray electrons. 
Figure~\ref{fig:gspec} shows the observed spectrum of atmospheric gamma rays 
at 7.4~g~cm$^{-2}$, 
which is well represented by 
\begin{equation}
J_{\gamma} = (3.19{\pm}0.59){\times}10^{-4} (\frac{E}{\rm 100GeV})^{-2.77{\pm}0.21} 
({\rm m^{-2} s^{-1} sr^{-1} GeV^{-1}}). 
\label{eq:gamma_spec}
\end{equation}
The spectral shape is in good agreement with the ECC and 
the calculated results from hadronic interactions of primary cosmic rays 
with the overlying atmosphere \cite{yoshida06}. 
At the normalized depth of 7.4~g~cm$^{-2}$, 
the gamma-ray flux agrees with the BETS 
in the energy region of 10 -- 20~GeV \cite{kasahara02}, 
and is consistent with that of ECC 
within 95\% confidence level of statistical errors.

\section{Results and discussions}
\label{sec:result}

%
\subsection{Energy spectrum of electrons}
\begin{figure}
 \begin{center}
  \includegraphics[width=100mm]{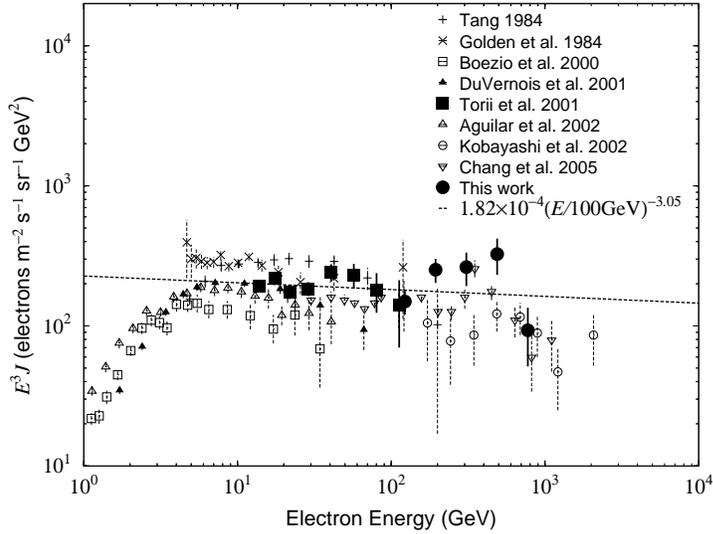}
 \end{center}
\caption{Electron energy spectrum observed with 
PPB-BETS (solid circles) 
in comparison with the energy spectra of 
BETS (solid squares) ant the other observations. 
The dash line shows the best fit power-law function of the combined spectrum 
of PPB-BETS and BETS with an index of $-3.05{\pm}0.05$.}
\label{fig:espec}
\end{figure}

Figure~\ref{fig:espec} presents the electron energy spectrum 
multiplied by the cube of energy, 
in comparison with other electron observations 
\cite{tang84, golden84, boezio00, duvernois01, torii01, aguilar02, 
kobayashi02, chang05}. 
Table~\ref{tab:elec_flux} presents the flux of electrons 
at the top of atmosphere in each energy interval. 
Our spectrum agrees well with the extrapolated spectrum of 
the BETS\cite{torii01} at 100~GeV. 
The combined spectrum of PPB-BETS and BETS can be 
represented by a single power-law function of 
\begin{equation}
J_{e} = (1.82{\pm}0.13){\times}10^{-4} (\frac{E}{\rm 100GeV})^{-3.05{\pm}0.05} 
({\rm m^{-2} s^{-1} sr^{-1} GeV^{-1}}) 
\label{eq:elec_spec}
\end{equation}
in the energy range of 10~GeV to 800~GeV. 
The reduced chi-square value is 1.605 for 11 degrees of freedom. 
A power-law spectrum is acceptable at the 95~\% confidence level 
of statistical errors.

The energy spectrum exceeding 100~GeV is crucial to detect the nearby 
SNRs as discussed by Kobayashi et al. (2004) \cite{kobayashi04}, 
and electron-positron pairs from 
Kaluza-Klein dark matter annihilations \cite{cheng02}. 
Although the data statistics of our results are insufficient to discuss 
the details of the contribution of nearby SNRs and/or WIMP dark matter, 
our energy spectrum may indicate a sign of a structure in the several 100~GeV region, 
as shown in Fig.~\ref{fig:espec}. 
Similar structure in the electron energy spectrum 
is reported by the ATIC-2 observations \cite{chang05}. 
Therefore, 
although the energy spectrum with PPB-BETS cannot reject a single power-law function, 
both the observations with PPB-BETS and ATIC-2 
may indicate a significant spectral structure. 
For the future observations, 
we are planning to increase greatly our statistics 
by experiments such as CALET 
with the geometrical factor of nearly 1~m$^{2}$sr 
and the observation time of 3 years 
on the International Space Station \cite{torii06}. 
It is expected that CALET can detect these distinctive signatures 
from the nearby SNRs and WIMP annihilations \cite{torii08}.

\subsection{Arrival directions of electrons}

\begin{figure}
 \begin{center}
   \includegraphics[width=100mm]{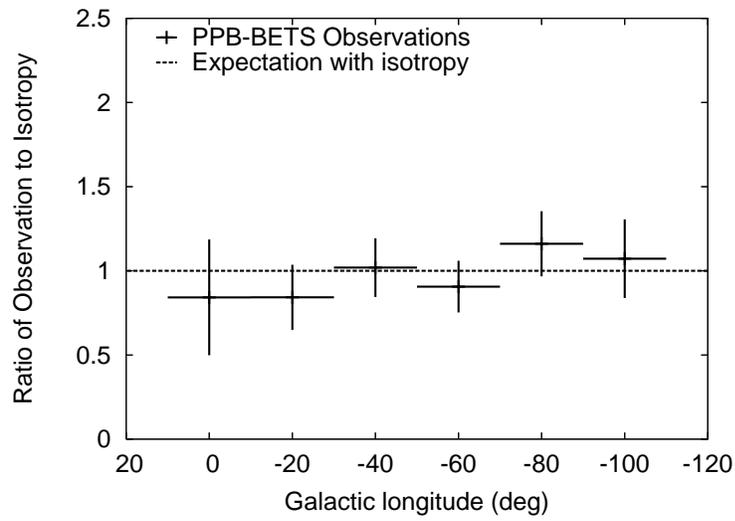}
 \end{center}
\caption{A ratio of the electron flux observed with PPB-BETS 
to the isotropic distribution  
along the Galactic longitude. }
\label{fig:obs2isotropy}
\end{figure}

Incident directions of electrons on the PPB-BETS detector 
are determined by using the shower axis 
with an accuracy of around 0.5~degree. 
Since the attitudes of PPB-BETS instrument are determined with 
sun aspect sensors and geomagnetic aspect sensors, 
we can determine arrival directions of cosmic-ray electrons 
with an accuracy of several degree. 
For the determination of attitudes with geomagnetic aspect sensors, 
we referred to the International Geomagnetic Reference Field (IGRF) model \cite{igrf05}. 
From these capabilities of PPB-BETS, 
for the first time we derived 
the arrival directions of high-energy electrons. 
Figure~\ref{fig:obs2isotropy} presents a ratio of the observed distribution 
above $\sim$100~GeV to the isotropic distribution along the Galactic longitude. 
As shown in Fig.~\ref{fig:obs2isotropy}, 
the arrival directions of electrons are consistent with the isotropic distribution.

Because the rate of energy loss due to the synchrotron and inverse Compton processes 
is much higher for electrons than for nuclei, 
the degree of anisotropy of high-energy cosmic ray electrons 
is expected to be higher than that of the nuclear component 
\cite{shen71, ptuskin95}. 
The magnitude of anisotropy 
is expected to be only around 1~\% in the 100~GeV -- 1~TeV 
in the direction of Vela at $(\ell, b) = (-96^{\circ}.1, -3^{\circ}.3)$ 
by using the calculated results by Kobayoashi et al. (2004) 
\cite{kobayashi04, yoshida08b}. 
The first result of the electron arrival directions by PPB-BETS 
shows no significant anisotropy within statistical errors. 
For the future observations, 
the detection of anisotropy toward the sources will enable us to 
identify the cosmic-ray electron sources, 
together with the unique signatures of the energy spectrum described above.

\section{Acknowledgment}

The PPB-BETS project was carried out in collaboration with NIPR and ISAS/JAXA. 
We express special thanks to the members who
participated in the PPB-BETS project, including the members in the JARE-44, 
who carried out the excellent and successful balloon flight at Syowa Station. 
We also thank the staffs of the H4 beam line of CERN-SPS 
for their kind support. 
This work was partly supported by Grants in Aid for Scientific Research 
on Priority Area A (Grant No.14039212) and 
Scientific Research C (Grant No.16540268, No.18540293).



\newpage

\begin{table}
\caption{Basic parameters of the PPB-BETS}
\label{tab:parameters}
\begin{center}
\begin{tabular}{lrl}
\hline
Instrument weight &  200~kg & including un-pressurized gondola \\
(Total weight     & 480~kg  & including ballast) \\ 
Power consumption & 70~W & supplied by solar batteries \\
Level altitude & $\sim$35~km & by automatic level control \\
Data transfer rate & 2.4~kbps & by the Iridium phone line  \\ 
                   & 64~kbps  & by the down link to the station \\
\hline
Energy range & 10~GeV--1~TeV & by two trigger modes \\
Effective geometrical factor & $\sim$300~cm$^2$sr & $>100$~GeV by simulation \\
Energy resolution & 10-20~\% & in 10~GeV--1~TeV by plastic scintillators \\
Angular resolution & 0.4--0.6~deg & by scintillating fibers \\
\hline
\end{tabular}
\end{center}
\end{table}

\begin{table}
\caption{Electron flux at the top of atmosphere}
\label{tab:elec_flux}
\begin{center}
\begin{tabular}{ccc}
\hline\hline
$E$           &  $\bar{E}$ & Intensity$^{a}$  \\
(GeV)         &  (GeV) & $({\rm m^{-2} s^{-1} sr^{-1} GeV^{-1}})$ \\
\hline
100.0--158.5  & 122.6  & $(8.1{\pm}1.6){\times}10^{-5}$ \\
158.5--251.2  & 194.4  & $(3.4{\pm}0.7){\times}10^{-5}$ \\
251.2--398.1  & 308.0  & $(9.0{\pm}2.4){\times}10^{-6}$ \\
398.1--631.0  & 488.2  & $(2.8{\pm}0.8){\times}10^{-6}$ \\
631.0--1000.0 & 773.7  & $(2.0{\pm}0.9){\times}10^{-7}$ \\
\hline
\multicolumn{3}{l}{$^{a}$ The quoted errors are the statistical error.}
\end{tabular}
\end{center}
\end{table}

\newpage

\end{document}